\documentclass{article}

\usepackage{arxiv}

\usepackage[utf8]{inputenc} 
\usepackage[T1]{fontenc}    
\usepackage{hyperref}       
\usepackage{url}            
\usepackage{abstract}
\usepackage{booktabs}       
\usepackage{amsfonts}       
\usepackage{nicefrac}       
\usepackage{microtype}      
\usepackage{cleveref}       
\usepackage{lipsum}         
\usepackage{graphicx} 
\usepackage{natbib}
\usepackage{doi}

\title{How does online teamwork change student \\ communication patterns in programming courses?}

\date{March 31, 2022}

\author{ \href{https://orcid.org/0000-0002-7476-5749}{\includegraphics[scale=0.06]{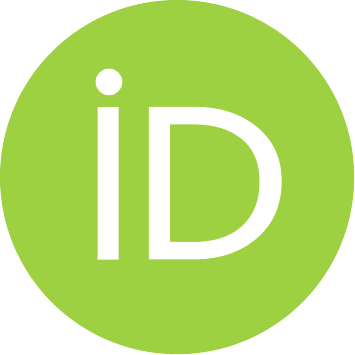}\hspace{1mm}N. Kozhevnikova} \\
	Department of Computer Science\\
	Perm State University\\
}
\graphicspath{ {images/} }

\hypersetup{
pdftitle={How does online teamwork change student communication patterns in programming courses},
pdfauthor={N. Kozhevnikova},
pdfkeywords={MOOC, blended learning, communication pattern},
}

\begin{document}
\maketitle

\begin{abstract}

Online teaching has become a new reality due to the COVID-19 pandemic raising a lot of questions about its learning outcomes. Recent studies have shown that peer communication positively affects learning outcomes of online teaching. However, it is not clear how collaborative programming tasks change peer communication patterns in the learning process. 
\par
In this study, we compare communication patterns in MOOCs where peer communication is limited with those of a blended course in which students are involved in online peer instruction. We used a mixed-method approach comprising automated text analysis and community extraction with further qualitative analysis.
\par
The results show that students prefer to seek help in programming from peers and not the teacher. Team assignment helped to support this habit. Students communicated more positively and intensively with each other, while only team leaders communicated with the instructor reducing teacher overload. This shift could explain how peer communication improves learning outcomes, as has been shown in previous studies on MOOCs.

\end{abstract}


\section{Introduction}

Recently there has been renewed interest in online learning. COVID-19 has changed the way we teach, thus raising questions about how the instructional design of online courses affects learning outcomes. Some universities turned to massive open online courses (MOOCs) to provide knowledge during the lockdown. Some teachers have also developed a hybrid approach combining online activities with offline classes when public health restrictions are milder. These attempts raised a lot of questions about the effects of online teaching on student learning outcomes \citep{crick_impact_2020,su_factors_2021,wang_analyzing_2021}.
\par
MOOCs can be classified based on the role of learners in the learning process \citep{hew_students_2014}. Extended Massive Open Online Courses (xMOOCs) mostly follow instructivist approach: learning objectives are predefined by the teacher, peer communication opportunities are limited to forum discussions and peer review assignments \citep{fidalgo-blanco_massive_2016, yousef_state_2015}. As opposed to xMOOCs, connectivist massive open online courses (cMOOCs) follow the connectivist approach: students have an active role in the development of course materials and are involved in open communication, including networking outside the course. Despite such problems as high drop rates and lack of student engagement \citep{zheng_understanding_2015}, xMOOCs are now the most common type of courses, since this approach is supported by leading platforms like Coursera, Edx or Udemy.
\par
Students can participate in MOOCs in self-paced or instructor-paced settings. Instructor-paced courses, also known as cohort MOOCs, have limited enrollment time and restricted intermediate and final deadlines. Self-paced courses provide students with an opportunity to join any time, therefore they have no other students learning with them. Previous studies have shown that self-paced courses require high self-regulation skills \citep{kizilcec_self-regulated_2017}, which in many cases are underdeveloped. 
\par
According to the Community of Inquiry (CoI) theoretical framework \citep{garrison_first_2010}, high quality education is driven by teacher presence, cognitive presence and social presence. Social presence requires rich peer interactions that could be hard to achieve online, especially in self-paced courses. Cognitive presence is a construction of meaning which is built by challenges, questions and exploration of problems. Teacher presence includes course organisation and management, mediation of social relationships, building understanding and student motivation. 
\citet*{moore_setting_2019} showed that lack of cohort does not significantly affect cognitive processing. It is course instruction design that has a stronger effect. Therefore, student-student interaction should be supported by the instructional design of MOOCs.
\par
Blended learning is an effort to combine the advantages of online learning while avoiding the challenges that MOOCs face \citep{anthony_blended_2020,ashraf_systematic_2021,rasheed_challenges_2020}. There is no established definition of blended learning but it is usually used to refer to some combination of online learning with offline classes and activities. One popular approach to blended learning is flipped classroom. In flipped classroom, students study theory in an online environment followed by an offline class where they can deepen their understanding with the help of the teacher \citep{alammary_blended_2014,anthony_blended_2020,tullis_why_2020}. Flipped classroom fits well with the active learning model, since it spares the in-class time for intensive practise. Such face-to-face lessons are often conducted using peer-learning activities in which students help each other to solve case studies or practical tasks, thus teaching each other. \citet{freeman_active_2014} showed that active learning improves student results in STEM disciplines. Although beneficial, blended learning creates challenges for teachers and students. Students need to have self-regulation skills; teachers are often required to do more work on content creation and course management, and both have to be proficient with the technology \citep{brown_blended_2016,rasheed_challenges_2020}.
\par
Another approach to blended learning is practice in a virtual environment which is common in teaching medical students. Due to the nature of tasks, it is also extensively used to teach computer sciences. However, students often work on their online assignments independently, receiving little benefit from peer-learning.
\par
Current research shows that students have better results when working on programming assignments collaboratively. Pair programming is one of the approaches to collaborative software development. There are several studies on pair programming in education showing that this technique increases learning outcomes, especially for women \citep{hanks_pair_2011,salleh_empirical_2011,umapathy_meta-analysis_2017}. Peer-leading approach with small teams in STEM disciplines also shows positive impact on learning outcomes \citep{herman_comparison_2020-1,wilson_small_2016}. Another collaborative programming approach is introducing peer code review in programming courses. According to \citep{lin_using_2021} code review in educational practice improves understanding of computational concepts and increases learning engagement and satisfaction.
\par
Since industrial software development includes collaborative programming as well as peer code review and peer leadership, we supposed that programming disciplines could use IT-industry teamwork practices to improve results of online courses. The IT industry has well established approaches to remote software development. Existing online services help to manage teamwork and software development. Therefore we designed a blended learning course that utilises those practises in an educational environment. To manage student teamwork, we used services that are common for the industry: Trello, Github and Discord. Current experience of using Github in education shows that students find it convenient for collaborative work and feel more prepared for the future \citep{fiksel_using_2019,hsing_using_2019}. The use of Trello and Discord helps to decrease the amount of managerial work for the teacher while teaching students state-of-the-art industry tools.
\par
Student feedback on the collaborative software development course was positive, therefore we wondered if a similar design can be used to overcome the problems of instructor-paced MOOCs discussed above. In order to answer this question, we explored how online teamwork development affected peer communication patterns in both instructional designs. Therefore, in this study we investigate how communication patterns differ between traditional xMOOCs and the aforementioned  blended course with collaborative project work.
\par
This study uses a mixed-method approach. The data were collected from forums of two MOOCs as well as from text messages and comments from Github, Trello and Discord used in the blended collaborative development course. We merged, preprocessed the communication data and applied topic modeling, sentiment analysis and community extraction in a communication network. Finally, we conducted qualitative analysis of extracted communication patterns and determined how they varied depending on the instructional design. 
\par
Our results showed that collaborative work changes the way students communicate. In both approaches, students prefer to discuss programming with peers, not with the instructors. In addition, it has been observed that when students work in a team they can have more intensive and positive discussions of their tasks. There is also a tendency for students to divide into leaders and those who refrain from active participation. Leaders take on the responsibility of teacher-student communication providing the team with all the information and supporting those who lack self-regulation skills.
\par
This knowledge could help understand how peer communication positively affects learning outcomes in MOOCs. It might also help teachers understand how their instructional choices might affect social and teacher presence in the course.
\par
Studies of non-English educational data are quite rare due to a smaller number of open online courses on other languages and therefore a relatively limited amount of data available. However, cultural and organisational differences could affect the learning process and student communication. Therefore, we tried to investigate whether our educational data support the state-of-the-art research on English courses.
\par
In terms of methodology, this study revealed one particular limitation in state-of-the-art automated text analyses. Students of programming disciplines use a mix of their native language, English and programming languages in their discussions. We were unable to find research on code mixing of this type. Also there is still little progress being made in the field of extracting knowledge from mixed natural and programming languages. This limitation raises problems with automated analysis of student questions in programming courses that are important for question answering educational bots. In this work, we decided to skip information that is expressed in different languages and conducted analyses for the primary (Russian) language only. 
Currently, steps are being made towards automating peer learning in an online environment. We showed that collaborative work similar to IT industry practises changes communication in a positive direction. The next step is to try and implement instructional designs which involve active peer-communication and peer-instruction in MOOCs. However, teamwork inside a MOOC can raise a bunch of new questions and require further investigation.
\par
This paper is structured into four main sections. In the first part, we discuss the data and methods that were used to conduct the analyses. The second part describes our results for MOOCs and the blended collaborative development course. In the third part of the paper, we discuss the main findings and limitations of the current work. We conclude by outlining possible directions for further research.

\section{Material and methods}

\subsection{Data collection}
Data for this study were collected from two “Open Education” MOOCs. “Open Education” is a branded instance of Open edX hosted by local Open Education association. The “Web development” and “Advanced web development” courses delivered via the platform are provided by ITMO University. These are instructor-paced courses. Students enrol twice a year and study in a cohort. The courses teach Python and its application to web development. They have a structure traditional for xMOOCs when video lectures are followed by tests and programming projects. The courses also offer exams for those who choose to pay for a certificate. For the “Web development” course, the data cover the period from autumn 2017 to spring 2020. As the “Advanced web development” course was released later, the respective data cover a shorter period: from autumn 2019 to spring 2020.
\par
The third dataset comes from the “Team software development methodology” course that we designed with a Community of Inquiry framework in mind. It aims to support collaborative learning while students study in a blended environment. Students engaged with learning materials in Moodle, discussed cases in offline lessons and worked on their own project with the help of real-world team development tools. Students were divided into several teams of 4-6 people, developed their application idea and created functioning software. Students used Discord for communication with peers and the tutor, Github for collaborative code development and Trello for agile management. Real-world services used in this course provide an opportunity to download activity data using their API. Following ethical guidelines, we received student consent to use the data for research. These data were downloaded, combined, cleaned and anonymised for further analysis. Dialogue data were collected from Discord team chats, comments from agile boards on Trello and from GitHub discussions and pull requests. 
\par
Although the MOOCs and the collaborative blended course teach different subjects, we believe that they have a lot in common to make such a comparison valid. They both require programming skills. Though the “Team software development methods” course does not strictly limit the choice of a development technology, students often select web or mobile-based projects. We would like to note that students rarely have basic knowledge of web development by the start of the “Team development methodology” course and many of them decide to develop two sets of skills in one project. Therefore, we believe that students of the selected MOOCs and the collaborative blended course have similar prior knowledge and learning needs which justifies the comparison of communication patterns in these courses.

\subsection {Data analysis}
In order to reveal how teamwork on a real-world project changes communication patterns in an online environment, we used mixed-methods analysis. We combined NLP (Natural language processing) analysis with a learning analytics approach. \par
First, MOOC dialogue texts, Discord chat logs and comments from agile board cards were preprocessed. Texts were tokenized, lemmatized and lowercased; stop words and punctuation were removed. \par
Since students on programming courses tend to share a lot of code in forums, this code can occur in any part of their message. It could be a function name from the assignment or lecture, some code they experience difficulties with, or a sample code improvement proposal. Messages with a combination of languages could provide some meaningful insights into communication patterns, however it is hard to detect the language involved \citep{jain_language_2014}. This is even more true for a mix of English and programming languages. A search of the literature revealed few studies on code-mixed extraction in the case of mixing programming languages \citep{jose_survey_2020}. Current research mostly focuses on natural language mixing, but we could not find a Russian-English mixed code dataset. Due to these limitations of code mixed topic modeling, we skipped programming code or mixed English words and processed only nouns, adjectives, verbs and adverbs in the primary language.
\par
To extract topics in forum messages, we applied Latent Dirichlet Allocation (LDA) from Gensim python library. Although LDA is often used for long texts and could produce worse results on texts shorter than 20 words \citep{yan_biterm_2013}, it still proves to be helpful in analysing (on) short texts  \citep{albalawi_using_2020}. \par
The LDA algorithm requires a predefined number of topics. Previous research has shown that selection of the number of topics is crucial for short-text data, since If the number of topics is too small, the topics are too general, and if this number is too large, topics will overlap \citep{ottesen_steinskog_a_twitter_2017}. To overcome these problems, BTM algorithm was created for short texts \citep{cheng_btm_2014}. However the BTM algorithm is less popular and therefore its metrics require further investigation. In our study we used the BTM package for R Studio.
\par
It is common to pick a number of topics using grid search with the coherence intrinsic metric, because it leads to better human interpretability \citep {roder_exploring_2015, syed_full-text_2017}, however, in our case high coherence was misleading. This occured due to the length of the texts and a small amount of messages in the blended collaborative course. In addition, students on online learning courses mostly discuss the content and assignments which are hard to distinguish for topic modelling algorithms. Therefore, we used both LDA and BTM models for topic modelling and then visually inspected topics where models showed controversial results. 
\par
The next step was sentiment analysis with the Dostoevsky python library. The Dostoevsky model is trained on the RuSentiment dataset and returns a sentiment classification of ‘positive’, ‘negative’, ‘neutral’, ‘speech’ as well as ‘skip’ for unclear cases. Sentiment scores for each class were also added to forum data. A message was said to have some tonality if the value for this tonality was higher than 0.7.
\par
Finally, we conducted community structure analysis using an algorithm proposed by \citep{traag_louvain_2019}. The student communication network of MOOCs was built based on the following principles: a response in a forum became an edge between the response author and the post author; a comment became an edge between the comment author and the comment recipient; a response on course-wide posts created by the instructor was not considered as an edge until someone answered it specifically.
\par
The communication network of the blended collaborative course includes messages from all the used services. Comments on tasks from the project management services (Trello) resulted in edges between a task assignee, or a mentioned person and a commenter. Pull request comments in Github repositories resulted in edges between a pull request owner or an assignee, if exists, or a mentioned person and a commenter.
\par
Students tend to communicate in Discord with little to no direct answering or mentions. We created an edge in the communication network for a direct answer or a mention. Mentions like ‘everyone’ or ‘here’ were not considered an edge until some further response. Messages that followed each other in less than one minute were attributed to the same conversation. In this case students probably answered to messages that had been sent right before. Messages sent with a longer delay were considered direct communication based on mentions or direct answers only.
\par
After detecting learning communities, we manually inspected  the discovered topics, tonality and networks and labelled the most frequent patterns for the traditional MOOC, blended MOOC and collaborative blended course. We investigated how student-student and student-teacher communities differed. We also reviewed changes in topics and sentiment inside different communities. 

\section{Results}
After preprocessing and removing duplicates, the MOOCs forum data contained 21378 text messages and the collaborative blended course data contained 924 messages.
\subsection{MOOCs patterns}
The best coherence metric values for xMOOCs topic modeling were achieved with 5 or 8 topics. However, manual review showed that this number of topics leads to an overlap and the results are hard to interpret. Only three of the topics were distinguishable enough to assign them some label.

\begin{figure}[t]
	\centering
	\includegraphics[width=0.7\textwidth]{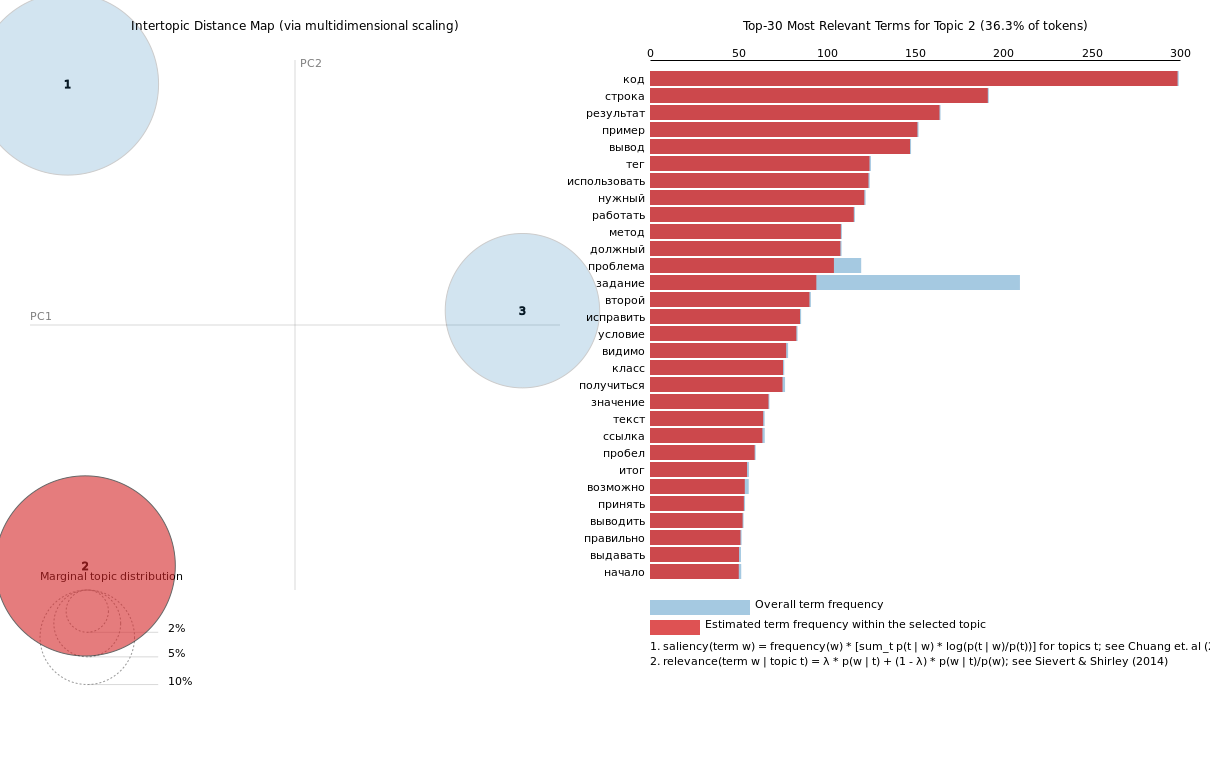}
	\caption{Visualization of a topic and top words: Programming.}
	\label{fig:fig1}
\end{figure}

Topics extracted from the MOOC forum texts were reviewed and labelled manually as course management issues, programming and learning material comprehension. Programming is very code specific and its keywords include such terms as ‘function’, ‘method’, ‘class’, ‘list’, ‘print’ (see Figure \ref{fig:fig1}). In these messages, students share specific details about difficulties in programming they encounter. Learning material comprehension, similar to Programming as it may seem, is less about programming and more about understanding lectures and analysing mistakes in course assignments. Course management issues are about certificates, deadlines and various technical problems. The topic keywords were added to forum data for learning analytics.
\par
\begin{figure}[ht]
	\centering
	\includegraphics[width=0.5\textwidth]{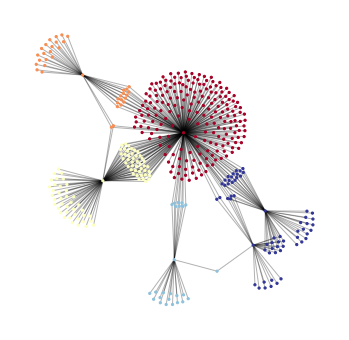}
	\caption{Teacher-student communication in MOOC.}
	\label{fig:fig2}
\end{figure}

Communities extracted from the data show that students tended to communicate only with one tutor (see Figure  \ref{fig:fig2}). 
\par

\begin{figure}[t]
	\centering
	\includegraphics[width=0.5\textwidth]{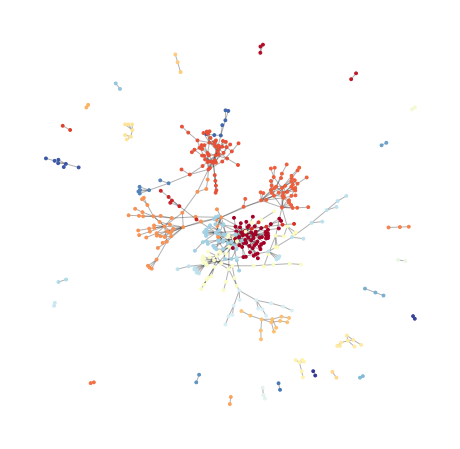}
	\caption{Student communication network in MOOCs.}
	\label{fig:fig3}
\end{figure}

Figure  \ref{fig:fig3} shows extracted communities in MOOCs forums. As can be seen that within one community, students interacted more actively while intercommunity communication was very low. It is important to note that student-teacher communication is very intensive and communities are built around the teachers.

\par

We also discovered that students discussed different topics with teachers and with each other. Thus, in our data only 3650 messages were labelled Programming in teacher-student communication, while 6231 messages were labelled Programming in student-student communication.  Learning material comprehension and course management issues, however, were discussed to a similar extent, as shown in Table \ref{tab:table1} .

\begin{table}[t]
	\caption{Topics in MOOC by student-teacher and student-student interaction (Russian keywords are translated)}
	\centering
	\begin{tabular}{|p{3cm}||p{3cm}|p{3cm}|p{4cm}|}
		\toprule
		Topic     & Student-Teacher	&Student-Student	&Keywords    \\
		\midrule
		Programming & 3650  & 6231   &task, code, result, problem, use, program, work, do, practice\\
		Learning comprehension     & 2727 & 2725   &error, line, example, do, understand, output, function, need, tag, method\\
		Course management     & 2617       & 3428  &course, question, week, may, why, test, answer, opportunity, version\\
		\bottomrule
	\end{tabular}
	\label{tab:table1}
\end{table}
\par
The analysis revealed that there were several very active communities of students who created most of the messages, while most of the students communicated much less or used passive communication patterns like upvoting messages.
Tonality of the messages was mostly neutral, however messages in student-teacher interactions were more likely to have a negative sentiment. It is also quite rare to see positive messages.  The proportion of positive messages to negative messages in MOOCs was 0.4.

\subsection{Collaborative blended course patterns}
\begin{figure}[t]
	\centering
	\includegraphics[width=0.7\textwidth]{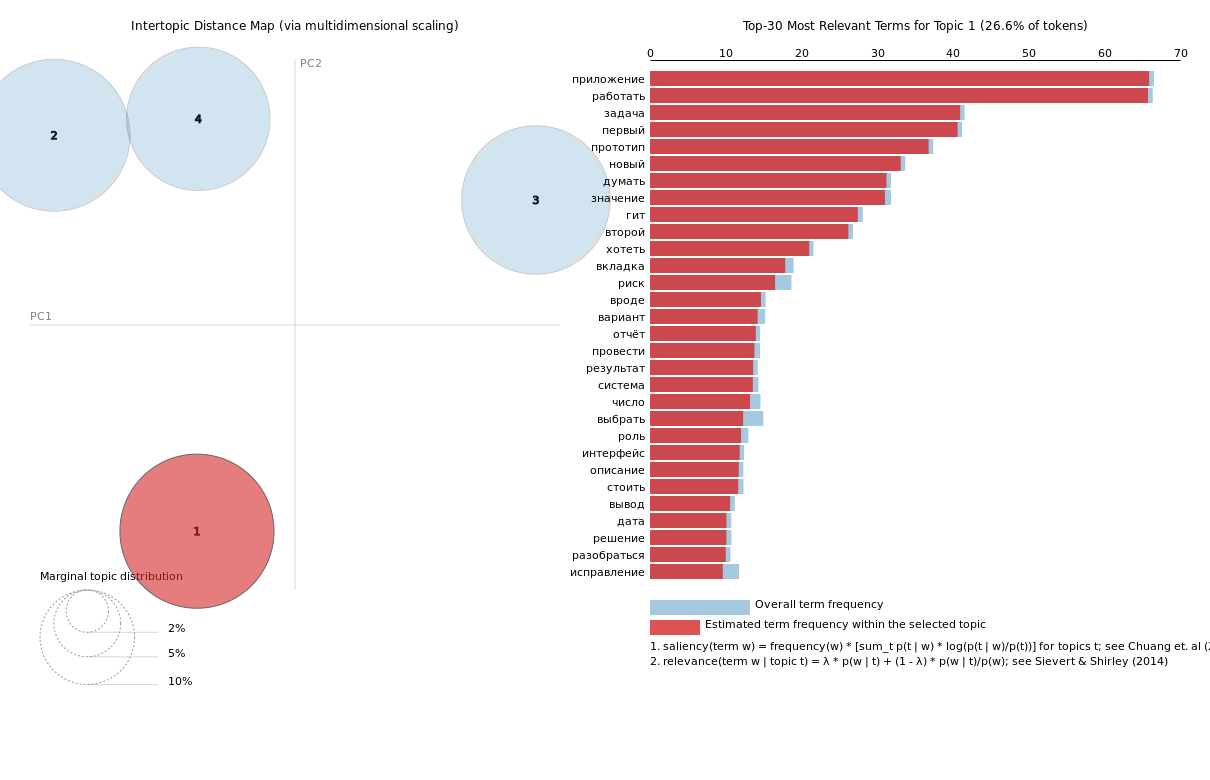}
	\caption{The course management topic in the collaborative blended course.}
	\label{fig:fig4}
\end{figure}
Topics extracted from the collaborative course data were diverse. Students tended to communicate about their collaboration, technical problems and application ideas. They also asked the tutor about assignment details. Due to the small amount of messages, automated algorithms had less agreement, therefore, a lot of messages were labeled manually.
We selected four main topics in the data: course management, product management, meetups and programming. Unlike in MOOCs data, course management questions were less about problems and more about assignment requirements (Figure \ref{fig:fig4}). Students asked about the final presentation of the project and how their decisions would affect the final mark. Product management topic is about generating application ideas and evaluating these ideas against the available time and resources. Since this course had an offline part, students also discussed their online and offline meetings. Those discussions were labeled Meetups. Finally, all the situations that were connected to programming received the label Programming, respectively.
\par

\begin{figure}[b]
	\centering
	\includegraphics[width=0.5\textwidth]{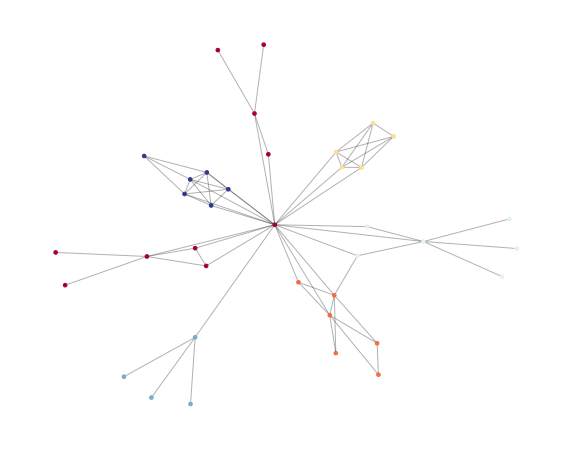}
	\caption{Communication within collaborative blended course.}
	\label{fig:fig5}
\end{figure}
It is no surprise that communities extracted from discussions were nearly the same as student teams. However, the structure of communication with the tutor was different from MOOCs. Students communicated using some hierarchy, where one or two leaders talked to the teacher and then discussed information inside their team (Figure \ref{fig:fig5}).
\par
With the teacher, students discussed such topics as course management and meetups topics (Table \ref{tab:table2}). The content of discussion varied according to the interlocutor: for example, within the topic of meetups, when addressing the teacher, students inquired about offline lessons while online meetups were discussed in student-student interactions. 

\begin{table}[t]
	\caption{Topics of blended course by student-teacher and student-student interactions (Russian keywords are translated)}
	\centering
	\begin{tabular}{|p{3cm}||p{3cm}|p{3cm}|p{4cm}|}
		\toprule
		Topic     & Student-Teacher	&Student-Student	&Keywords    \\
		\midrule
		Course management &73  & 212   &application, work, task, first, prototype\\
		Product management     &47 & 159   &do, template, team, model, page, button\\
		Meetups     & 56       &196  &course, question, week, may, why, test, answer, opportunity, version\\
		Programming &33	&195	&add, file, create, work, test, environment variables\\
		\bottomrule
	\end{tabular}
	\label{tab:table2}
\end{table}

Tonality of the communication was mostly neutral. However, students communicated in a more positive manner. The proportion of positive messages to negative messages in the collaborative blended course was 4. 

\section{Discussion}

The results show that differences in instructional design significantly affect communication patterns. In a collaborative environment, students build strong relationships inside the team and communicate with peers more than with a teacher. Peer-support affects their attitude, as changes from neutral-or-negative to neutral-or-positive message sentiment show.
\par
The intensity of interaction revealed by our mixed-method approach makes it quite clear that collaboration reduces the amount of work for the teacher. We might assume that students with low self-regulation skills rely on their peers with highly developed self-regulation skills. Thus, they receive all the information from their team leader reducing the number of messages sent to the teacher inquiring about similar problems. A closer look at the topic of meetups also shows that students pay particular attention to time management and this could help students who struggle to plan their work without external support. \par
From the results of this research, we can see that students tend to discuss programming tasks with each other and not with a teacher. This is true with the MOOCs data as well as with the blended collaborative course data. This could explain how intensive peer-communication increases learning outcomes. One reason might be that students consider the teacher being less available for discussions than their peers. Another reason might be that some students are rather shy to ask the teacher while communication with a peer could seem less stressful. Therefore, we might conclude that peer communication, indeed, has a direct influence on the development of programming skills. \par
However, the above results could be affected by several limitations. The main limitation of this study is the size of the datasets. Even though the MOOCs dataset covered several years, peer-communication was not very intensive and student-teacher communication dominated.  The limited amount of text data on the collaborative learning course, in its turn, can be attributed to a small number of students. Thus, the data were unbalanced which could have affected the results. Incorporating student-content clickstream data could help to overcome this imbalance and study communication patterns in more detail.\par
Another source of uncertainty arises from the decision to skip programming code that students post in their questions. In programming disciplines, code brings a lot of meaning and natural language just helps to clarify the student inquiry. However, we were unable to find a reliable solution for topic modelling of code mixing of natural and programming languages. Therefore our topics are very general combining all the programming problems in one cluster. Despite this limitation, we believe that this did not affect our results since topic modelling was used only as a part of manual qualitative analysis.\par
Current research on code mixing is mostly devoted to mixing two natural languages \citep{jose_survey_2020}. Some studies aim to support developers in information extraction \citep{palomba_automatic_2018,rodriguez_comparison_2019}, but most of them rely on tagged data like questions from  StackOverflow or compare the code only . There are also some attempts to use topic modeling in bug reports to help find bugs in the code \citep{lam_bug_2017,wang_bug_2018,zhou_where_2012}, but these results cannot be applied in a question answering context.\par
Mixing programming code with natural languages is common in developer communication in messengers or e-mails. Therefore, future work on topic modeling of the source code in the context of question answering is required to better understand how students communicate about programming tasks.\par
A natural progression of this work is to find out how communication patterns might change if more peer communication is introduced to MOOCs through collaborative assignments. This could raise new research questions. First, students in MOOCs often select this type of learning for the opportunity to study independently, therefore, a need to rely on a peer to complete assignments might cause frustration and even lead to a dropout. However, there is evidence that the number of students who can actually work independently and successfully finish the course is quite small \citep{reich_mooc_2019}. Therefore, despite some independent students might drop out, there is a reason to believe that overall learning outcomes and student engagement might improve. This is especially true for blended MOOCs that are used as an addition to offline academic courses.\par
A shift to collaborative programming assignments could also require more management from the teacher, thus creating overload. Even though our study showed that the number of questions to teachers decreases, it did not measure the amount of teacher management  needed to facilitate the conversations within Trello and Github. We suppose that some of those tasks could be automated by an educational bot. Thus, the bot could answer some common organisational questions, provide students with online tutorials on tools and automate some project management tasks. The development of such educational bots might also partially rely on the results of this study and other research similar in nature.

 \section{Conclusion}

Latest changes in higher education due to the pandemic will affect the way we teach for a long time. Even when there are no strong restrictions, lessons learnt from online teaching will keep changing the educational landscape. In this study we investigated how different online instructional designs affect the way students communicate with each other and with a teacher.\par
We analysed two approaches to teaching online: instructor-paced MOOCs and a blended course with a collaborative team assignment. We used MOOCs forum data and log data from team services that were used in the blended collaborative course. After preprocessing the text, we conducted topic modeling and sentiment analysis. We also extracted communities from communication networks.\par
The results show that incorporation of teamwork helps to restructure communication patterns within a course. While the number of questions to the teacher reduces, the number of peer-to-peer interactions increases. Students start to communicate in a more positive manner and help each other to manage time and tasks.\par
These findings can help teachers to choose an appropriate instructional design for their courses. For example, online project teamwork could become a good solution to the problems of instructor-paced MOOCs and require further research.

\section*{Acknowledgements}
The author would like to humbly thank Prof. Dmitriy Shtennikov from ITMO University for sharing the data from MOOC courses and his helpful contribution to data collection and analysis.

\bibliographystyle{authordate1}
\bibliography{article.bib}  

\begin{thebibliography}{}

\bibitem[\protect\citename{Alammary {\em et~al.}, }2014]{alammary_blended_2014}
Alammary, Ali, Sheard, Judy, \& Carbone, Angela. 2014.
\newblock Blended learning in higher education: {Three} different design
  approaches.
\newblock {\em Australasian Journal of Educational Technology}, {\bf 30}(4).

\bibitem[\protect\citename{Albalawi {\em et~al.}, }2020]{albalawi_using_2020}
Albalawi, Rania, Yeap, Tet~Hin, \& Benyoucef, Morad. 2020.
\newblock Using {Topic} {Modeling} {Methods} for {Short}-{Text} {Data}: {A}
  {Comparative} {Analysis}.
\newblock {\em Frontiers in Artificial Intelligence}, {\bf 3}(July), 42.

\bibitem[\protect\citename{Anthony {\em et~al.}, }2020]{anthony_blended_2020}
Anthony, Bokolo, Kamaludin, Adzhar, Romli, Awanis, Raffei, Anis Farihan~Mat,
  Phon, Danakorn Nincarean A. L.~Eh, Abdullah, Aziman, \& Ming, Gan~Leong.
  2020.
\newblock Blended {Learning} {Adoption} and {Implementation} in {Higher}
  {Education}: {A} {Theoretical} and {Systematic} {Review}.
\newblock {\em Technology, Knowledge and Learning}, Oct.

\bibitem[\protect\citename{Ashraf {\em et~al.}, }2021]{ashraf_systematic_2021}
Ashraf, Muhammad~Azeem, Yang, Meijia, Zhang, Yufeng, Denden, Mouna, Tlili,
  Ahmed, Liu, Jiayi, Huang, Ronghuai, \& Burgos, Daniel. 2021.
\newblock A {Systematic} {Review} of {Systematic} {Reviews} on {Blended}
  {Learning}: {Trends}, {Gaps} and {Future} {Directions}.
\newblock {\em Psychology Research and Behavior Management}, {\bf 14}(Oct.),
  1525--1541.

\bibitem[\protect\citename{Brown, }2016]{brown_blended_2016}
Brown, Michael~Geoffrey. 2016.
\newblock Blended instructional practice: {A} review of the empirical
  literature on instructors' adoption and use of online tools in face-to-face
  teaching.
\newblock {\em The Internet and Higher Education}, {\bf 31}(Oct.), 1--10.

\bibitem[\protect\citename{Cheng {\em et~al.}, }2014]{cheng_btm_2014}
Cheng, Xueqi, Yan, Xiaohui, Lan, Yanyan, \& Guo, Jiafeng. 2014.
\newblock {BTM}: {Topic} {Modeling} over {Short} {Texts}.
\newblock {\em IEEE Transactions on Knowledge and Data Engineering}, {\bf
  26}(12), 2928--2941.

\bibitem[\protect\citename{Crick {\em et~al.}, }2020]{crick_impact_2020}
Crick, Tom, Knight, Cathryn, Watermeyer, Richard, \& Goodall, Janet. 2020.
\newblock The {Impact} of {COVID}-19 and “{Emergency} {Remote} {Teaching}”
  on the {UK} {Computer} {Science} {Education} {Community}.
\newblock {\em Pages  31--37 of:} {\em United {Kingdom} \& {Ireland}
  {Computing} {Education} {Research} conference.}
\newblock Glasgow United Kingdom: ACM.

\bibitem[\protect\citename{Fidalgo-Blanco {\em et~al.},
  }2016]{fidalgo-blanco_massive_2016}
Fidalgo-Blanco, Ángel, Sein-Echaluce, María~Luisa, \& García-Peñalvo,
  Francisco~José. 2016.
\newblock From massive access to cooperation: lessons learned and proven
  results of a hybrid {xMOOC}/{cMOOC} pedagogical approach to {MOOCs}.
\newblock {\em International Journal of Educational Technology in Higher
  Education}, {\bf 13}(1), 24.

\bibitem[\protect\citename{Fiksel {\em et~al.}, }2019]{fiksel_using_2019}
Fiksel, Jacob, Jager, Leah~R., Hardin, Johanna~S., \& Taub, Margaret~A. 2019.
\newblock Using {GitHub} {Classroom} {To} {Teach} {Statistics}.
\newblock {\em Journal of Statistics Education}, {\bf 27}(2), 110--119.

\bibitem[\protect\citename{Freeman {\em et~al.}, }2014]{freeman_active_2014}
Freeman, S., Eddy, S.~L., McDonough, M., Smith, M.~K., Okoroafor, N., Jordt,
  H., \& Wenderoth, M.~P. 2014.
\newblock Active learning increases student performance in science,
  engineering, and mathematics.
\newblock {\em Proceedings of the National Academy of Sciences}, {\bf 111}(23),
  8410--8415.

\bibitem[\protect\citename{Garrison {\em et~al.}, }2010]{garrison_first_2010}
Garrison, D.~Randy, Anderson, Terry, \& Archer, Walter. 2010.
\newblock The first decade of the community of inquiry framework: {A}
  retrospective.
\newblock {\em The Internet and Higher Education}, {\bf 13}(1-2), 5--9.

\bibitem[\protect\citename{Hanks {\em et~al.}, }2011]{hanks_pair_2011}
Hanks, Brian, Fitzgerald, Sue, McCauley, Renée, Murphy, Laurie, \& Zander,
  Carol. 2011.
\newblock Pair programming in education: a literature review.
\newblock {\em Computer Science Education}, {\bf 21}(2), 135--173.

\bibitem[\protect\citename{Herman \& Azad, }2020]{herman_comparison_2020-1}
Herman, Geoffrey~L., \& Azad, Sushmita. 2020.
\newblock A {Comparison} of {Peer} {Instruction} and {Collaborative} {Problem}
  {Solving} in a {Computer} {Architecture} {Course}.
\newblock {\em Pages  461--467 of:} {\em Proceedings of the 51st {ACM}
  {Technical} {Symposium} on {Computer} {Science} {Education}}.
\newblock Portland OR USA: ACM.

\bibitem[\protect\citename{Hew \& Cheung, }2014]{hew_students_2014}
Hew, Khe~Foon, \& Cheung, Wing~Sum. 2014.
\newblock Students’ and instructors’ use of massive open online courses
  ({MOOCs}): {Motivations} and challenges.
\newblock {\em Educational Research Review}, {\bf 12}(June), 45--58.

\bibitem[\protect\citename{Hsing \& Gennarelli, }2019]{hsing_using_2019}
Hsing, Courtney, \& Gennarelli, Vanessa. 2019.
\newblock Using {GitHub} in the {Classroom} {Predicts} {Student} {Learning}
  {Outcomes} and {Classroom} {Experiences}: {Findings} from a {Survey} of
  {Students} and {Teachers}.
\newblock {\em Pages  672--678 of:} {\em Proceedings of the 50th {ACM}
  {Technical} {Symposium} on {Computer} {Science} {Education}}.
\newblock Minneapolis MN USA: ACM.

\bibitem[\protect\citename{Jain \& Bhat, }2014]{jain_language_2014}
Jain, Naman, \& Bhat, Riyaz~Ahmad. 2014.
\newblock Language identification in code-switching scenario.
\newblock {\em Pages  87--93 of:} {\em Proceedings of the {First} {Workshop} on
  {Computational} {Approaches} to {Code} {Switching}}.

\bibitem[\protect\citename{Jose {\em et~al.}, }2020]{jose_survey_2020}
Jose, Navya, Chakravarthi, Bharathi~Raja, Suryawanshi, Shardul, Sherly,
  Elizabeth, \& McCrae, John~P. 2020.
\newblock A {Survey} of {Current} {Datasets} for {Code}-{Switching} {Research}.
\newblock {\em Pages  136--141 of:} {\em 2020 6th {International} {Conference}
  on {Advanced} {Computing} and {Communication} {Systems} ({ICACCS})}.
\newblock Coimbatore, India: IEEE.

\bibitem[\protect\citename{Kizilcec {\em et~al.},
  }2017]{kizilcec_self-regulated_2017}
Kizilcec, René~F., Pérez-Sanagustín, Mar, \& Maldonado, Jorge~J. 2017.
\newblock Self-regulated learning strategies predict learner behavior and goal
  attainment in {Massive} {Open} {Online} {Courses}.
\newblock {\em Computers \& Education}, {\bf 104}(Jan.), 18--33.

\bibitem[\protect\citename{Lam {\em et~al.}, }2017]{lam_bug_2017}
Lam, An~Ngoc, Nguyen, Anh~Tuan, Nguyen, Hoan~Anh, \& Nguyen, Tien~N. 2017.
\newblock Bug {Localization} with {Combination} of {Deep} {Learning} and
  {Information} {Retrieval}.
\newblock {\em Pages  218--229 of:} {\em 2017 {IEEE}/{ACM} 25th {International}
  {Conference} on {Program} {Comprehension} ({ICPC})}.
\newblock Buenos Aires, Argentina: IEEE.

\bibitem[\protect\citename{Lin {\em et~al.}, }2021]{lin_using_2021}
Lin, Xuefen, Ma, Yanghui, Ma, Weifeng, Liu, Yang, \& Tang, Wei. 2021.
\newblock Using peer code review to improve computational thinking in a blended
  learning environment: {A} randomized control trial.
\newblock {\em Computer Applications in Engineering Education}, {\bf 29}(6),
  1825--1835.

\bibitem[\protect\citename{Moore {\em et~al.}, }2019]{moore_setting_2019}
Moore, Robert~L., Oliver, Kevin~M., \& Wang, Chuang. 2019.
\newblock Setting the pace: examining cognitive processing in {MOOC} discussion
  forums with automatic text analysis.
\newblock {\em Interactive Learning Environments}, {\bf 27}(5-6), 655--669.

\bibitem[\protect\citename{Ottesen~Steinskog {\em et~al.},
  }2017]{ottesen_steinskog_a_twitter_2017}
Ottesen~Steinskog, Asbj{\o}rn, Foyn~Therkelsen, Jonas, \& Gamb{\"a}ck,
  Bj{\"o}rn. 2017.
\newblock Twitter Topic Modeling by Tweet Aggregation.

\bibitem[\protect\citename{Palomba {\em et~al.}, }2018]{palomba_automatic_2018}
Palomba, Fabio, Zaidman, Andy, \& De~Lucia, Andrea. 2018.
\newblock Automatic {Test} {Smell} {Detection} {Using} {Information}
  {Retrieval} {Techniques}.
\newblock {\em Pages  311--322 of:} {\em 2018 {IEEE} {International}
  {Conference} on {Software} {Maintenance} and {Evolution} ({ICSME})}.
\newblock Madrid: IEEE.

\bibitem[\protect\citename{Rasheed {\em et~al.},
  }2020]{rasheed_challenges_2020}
Rasheed, Rasheed~Abubakar, Kamsin, Amirrudin, \& Abdullah, Nor~Aniza. 2020.
\newblock Challenges in the online component of blended learning: {A}
  systematic review.
\newblock {\em Computers \& Education}, {\bf 144}(Jan.), 103701.

\bibitem[\protect\citename{Röder {\em et~al.}, }2015]{roder_exploring_2015}
Röder, Michael, Both, Andreas, \& Hinneburg, Alexander. 2015.
\newblock Exploring the {Space} of {Topic} {Coherence} {Measures}.
\newblock {\em Pages  399--408 of:} {\em Proceedings of the {Eighth} {ACM}
  {International} {Conference} on {Web} {Search} and {Data} {Mining}}.
\newblock Shanghai China: ACM.

\bibitem[\protect\citename{Reich \& Ruipérez-Valiente, }2019]{reich_mooc_2019}
Reich, Justin, \& Ruipérez-Valiente, José~A. 2019.
\newblock The {MOOC} pivot.
\newblock {\em Science}, {\bf 363}(6423), 130--131.

\bibitem[\protect\citename{Rodriguez \& Carver,
  }2019]{rodriguez_comparison_2019}
Rodriguez, Danissa~V., \& Carver, Doris~L. 2019.
\newblock Comparison of {Information} {Retrieval} {Techniques} for
  {Traceability} {Link} {Recovery}.
\newblock {\em Pages  186--193 of:} {\em 2019 {IEEE} 2nd {International}
  {Conference} on {Information} and {Computer} {Technologies} ({ICICT})}.
\newblock Kahului, HI, USA: IEEE.

\bibitem[\protect\citename{Salleh {\em et~al.}, }2011]{salleh_empirical_2011}
Salleh, Norsaremah, Mendes, Emilia, \& Grundy, John. 2011.
\newblock Empirical {Studies} of {Pair} {Programming} for {CS}/{SE} {Teaching}
  in {Higher} {Education}: {A} {Systematic} {Literature} {Review}.
\newblock {\em IEEE Transactions on Software Engineering}, {\bf 37}(4),
  509--525.

\bibitem[\protect\citename{Su \& Guo, }2021]{su_factors_2021}
Su, Chien-Yuan, \& Guo, Yuqing. 2021.
\newblock Factors impacting university students' online learning experiences
  during the {COVID}-19 epidemic.
\newblock {\em Journal of Computer Assisted Learning}.
\newblock eprint: https://onlinelibrary.wiley.com/doi/pdf/10.1111/jcal.12555.

\bibitem[\protect\citename{Syed \& Spruit, }2017]{syed_full-text_2017}
Syed, Shaheen, \& Spruit, Marco. 2017.
\newblock Full-{Text} or {Abstract}? {Examining} {Topic} {Coherence} {Scores}
  {Using} {Latent} {Dirichlet} {Allocation}.
\newblock {\em Pages  165--174 of:} {\em 2017 {IEEE} {International}
  {Conference} on {Data} {Science} and {Advanced} {Analytics} ({DSAA})}.
\newblock Tokyo, Japan: IEEE.

\bibitem[\protect\citename{Traag {\em et~al.}, }2019]{traag_louvain_2019}
Traag, V.~A., Waltman, L., \& van Eck, N.~J. 2019.
\newblock From {Louvain} to {Leiden}: guaranteeing well-connected communities.
\newblock {\em Scientific Reports}, {\bf 9}(1), 5233.

\bibitem[\protect\citename{Tullis \& Goldstone, }2020]{tullis_why_2020}
Tullis, Jonathan~G., \& Goldstone, Robert~L. 2020.
\newblock Why does peer instruction benefit student learning?
\newblock {\em Cognitive Research: Principles and Implications}, {\bf 5}(1),
  15.

\bibitem[\protect\citename{Umapathy \& Ritzhaupt,
  }2017]{umapathy_meta-analysis_2017}
Umapathy, Karthikeyan, \& Ritzhaupt, Albert~D. 2017.
\newblock A {Meta}-{Analysis} of {Pair}-{Programming} in {Computer}
  {Programming} {Courses}: {Implications} for {Educational} {Practice}.
\newblock {\em ACM Transactions on Computing Education}, {\bf 17}(4), 1--13.

\bibitem[\protect\citename{Wang {\em et~al.}, }2021]{wang_analyzing_2021}
Wang, Xue, Lee, Youngjin, Lin, Lin, Mi, Ying, \& Yang, Tiantian. 2021.
\newblock Analyzing instructional design quality and students' reviews of 18
  courses out of the {Class} {Central} {Top} 20 {MOOCs} through systematic and
  sentiment analyses.
\newblock {\em The Internet and Higher Education}, {\bf 50}(June), 100810.

\bibitem[\protect\citename{Wang {\em et~al.}, }2018]{wang_bug_2018}
Wang, Yaojing, Yao, Yuan, Tong, Hanghang, Huo, Xuan, Li, Min, Xu, Feng, \& Lu,
  Jian. 2018.
\newblock Bug {Localization} via {Supervised} {Topic} {Modeling}.
\newblock {\em Pages  607--616 of:} {\em 2018 {IEEE} {International}
  {Conference} on {Data} {Mining} ({ICDM})}.
\newblock Singapore: IEEE.

\bibitem[\protect\citename{Wilson \& Varma-Nelson, }2016]{wilson_small_2016}
Wilson, Sarah~Beth, \& Varma-Nelson, Pratibha. 2016.
\newblock Small {Groups}, {Significant} {Impact}: {A} {Review} of {Peer}-{Led}
  {Team} {Learning} {Research} with {Implications} for {STEM} {Education}
  {Researchers} and {Faculty}.
\newblock {\em Journal of Chemical Education}, {\bf 93}(10), 1686--1702.

\bibitem[\protect\citename{Yan {\em et~al.}, }2013]{yan_biterm_2013}
Yan, Xiaohui, Guo, Jiafeng, Lan, Yanyan, \& Cheng, Xueqi. 2013.
\newblock A biterm topic model for short texts.
\newblock {\em Pages  1445--1456 of:} {\em Proceedings of the 22nd
  international conference on {World} {Wide} {Web} - {WWW} '13}.
\newblock Rio de Janeiro, Brazil: ACM Press.

\bibitem[\protect\citename{Yousef {\em et~al.}, }2015]{yousef_state_2015}
Yousef, Ahmed Mohamed~Fahmy, Chatti, Mohamed~Amine, Schroeder, Ulrik, Wosnitza,
  Marold, \& Jakobs, Harald. 2015.
\newblock The {State} of {MOOCs} from 2008 to 2014: {A} {Critical} {Analysis}
  and {Future} {Visions}.
\newblock {\em Pages  305--327 of:} {\em Computer {Supported} {Education}},
  vol. 510.
\newblock Cham: Springer International Publishing.
\newblock Series Title: Communications in Computer and Information Science.

\bibitem[\protect\citename{Zheng {\em et~al.}, }2015]{zheng_understanding_2015}
Zheng, Saijing, Rosson, Mary~Beth, Shih, Patrick~C., \& Carroll, John~M. 2015.
\newblock Understanding {Student} {Motivation}, {Behaviors} and {Perceptions}
  in {MOOCs}.
\newblock {\em Pages  1882--1895 of:} {\em Proceedings of the 18th {ACM}
  {Conference} on {Computer} {Supported} {Cooperative} {Work} \& {Social}
  {Computing}}.
\newblock Vancouver BC Canada: ACM.

\bibitem[\protect\citename{Zhou {\em et~al.}, }2012]{zhou_where_2012}
Zhou, Jian, Zhang, Hongyu, \& Lo, David. 2012.
\newblock Where should the bugs be fixed? {More} accurate information
  retrieval-based bug localization based on bug reports.
\newblock {\em Pages  14--24 of:} {\em 2012 34th {International} {Conference}
  on {Software} {Engineering} ({ICSE})}.
\newblock Zurich: IEEE.

\end{thebibliography}

\end{document}